\documentclass[a4paper,12pt]{article}
\usepackage{amssymb}

\usepackage{graphicx}
\usepackage{amsmath}
\usepackage{epsfig}

\input{tcilatex}

\begin{document}

\begin{flushright}
FTUV-02-0614 \\ 
\end{flushright}

%\title{\textbf{Unitarity Triangles and the Search for New Physics}}
\begin{center}
{\Large \textbf{Unitarity Triangles and the Search for New Physics}}\\

\medskip

F.J.Botella$^a$ \footnote{e-mail:francisco.j.botella@uv.es}\ ,   
G.C.Branco$^b$  \footnote{e-mail: gbranco@alfa.ist.utl.pt}\ ,  
M.Nebot$^a$  \footnote{e-mail:Miguel.Nebot@uv.es} \ and
M.N.Rebelo$^b$ \footnote{e-mail: rebelo@alfa.ist.utl.pt} 
%\maketitle

\bigskip

$^a$Departament de F\'\i sica Te\`orica and IFIC\\
Universitat de Val\`encia-CSIC, E-46100, Burjassot, Spain\\
$^b$Departamento de F\'\i sica \\
Instituto Superior T\'{e}cnico, P-1049-001,
Lisboa, Portugal\\
\end{center}
\begin{abstract}
Assuming that the Kobayashi-Maskawa mechanism gives the dominant
contribution to CP violation at low energies, 
we propose a novel way of testing the flavour sector of
the Standard Model which has the
potencial for discovering New Physics. Using $3 \times 3$ unitarity
of the $V_{CKM}$ matrix and choosing a complete set of rephasing 
invariant phases, we derive a set of exact relations in terms 
of measurable quantities, namely moduli of $V_{CKM}$ and arguments of 
rephasing invariant quartets. These tests complement the usual 
analysis in the $\rho$, $\eta$ plane and, if there is 
New Physics, may reveal its source.
\end{abstract}

\section{Introduction}

\noindent The advent of various B-factories has triggered an important
development in the study of CP violation, 
with both BaBar (SLAC) \cite{Aubert:2002gv} 
and Belle (KEK) \cite{Abe:2001xe} providing for the first
time evidence for CP violation outside the Kaon system. This new data and
its expected improvement in the near future 
\cite{Harrison:1998yr}, \cite{Ball:2000ba},
\cite{Hurth:2001yy}, \cite{Anikeev:2001rk},
will provide a stringent test of
one of the experimentally least constrained aspects of the Standard Model
(SM), namely the Kobayashi-Maskawa (KM) mechanism of CP violation.

So far, all experimental data on flavour physics and CP violation 
\cite{Branco:1999fs}, \cite{Bigi:yz}  are in
agreement with the SM \ and its KM mechanism.
This agreement is remarkable,
since one has to account for a large number of data with a small number of
parameters. The Cabibbo, Kobayashi and Maskawa (CKM) matrix is characterized
by four parameters which one can choose to be three angles $\theta _{i}$ and
the phase $\delta $ of the standard parametrization \cite{Groom:in}. 
The values of $s_{1}$, 
$s_{2}$ and $s_{3}$ ( $s_{i}=\sin \theta _{i}$ ) can be determined by the
experimental value of $\left| V_{us}\right| $, $\left| V_{cb}\right| $ 
and $\left| V_{ub}\right| $. Once these parameters are fixed, one has to fit,
using only the phase $\delta $, a large amount of data, 
including $\varepsilon _{K}$, $\varepsilon ^{\prime }/\varepsilon$, 
$\sin \left( 2\beta \right)$, $\Delta M_{B_{d}}$, 
$\Delta M_{B_{s}}$. It is remarkable that these
five experimental quantities can be fitted with only one 
parameter \cite{Hocker:2001xe}, namely
the KM phase $\delta $.

In this paper, we address the question of finding the best strategy to 
perform precision tests
of the SM mechanism of flavour mixing and CP violation,
while at the same time searching for the presence of New Physics. 

In view of the impressive success of the SM, one may wonder 
what is the motivation to look for Physics Beyond the SM. 
In what concerns CP violation, there are in
our opinion, two main motivations to look for New Physics 
and in particular new sources of CP violation:

\begin{itemize}
\item[(i)]  By now, it has been established that the strength of CP
violation in the SM is not sufficient to generate the observed Baryon
Asymmetry in the Universe (BAU), thus suggesting the need for new sources of
CP violation.

\item[(ii)]  Almost all extensions of the SM, including supersymmetric
extensions, have new sources of CP violation which can in principle be
detected at B-factories.
\end{itemize}

Throughout the paper, we will assume that the tree level weak decays
are dominated by the SM W-exchange diagrams, thus implying that the 
extraction of $\left| V_{us}\right| $, $\left| V_{ub}\right| $ 
and $\left| V_{cb}\right| $ from experiment continues to be valid
even in the presence of New Physics (NP). We will allow for 
contributions from NP in processes like 
$B_{d}^{0}-\overline{B}_{d}^{0}$ mixing and 
$B_{s}^{0}-\overline{B}_{s}^{0}$ mixing, as well as in penguin diagrams.
Since the SM contributes to these processes only at loop level, the
effects of NP are more likely to be detectable. Examples of processes
which are sensitive to NP, are the CP asymmetries corresponding to the
decays ${B^0}_d \rightarrow J/\Psi K_{s}$ and
${B^0}_d \rightarrow \pi^+ \pi^-$ which are affected by NP contributions
to $B_{d}^{0}-\overline{B}_{d}^{0}$ mixing. Significant contributions
to $B_{d}^{0}-\overline{B}_{d}^{0}$ and 
$B_{s}^{0}-\overline{B}_{s}^{0}$ mixing can arise in many of
the extensions of the SM, such as models with vector-like quarks
\cite{vlq1}, \cite{vlq2} and supersymmetric extensions 
of the SM \cite{susy}. Vector-like quarks naturally 
arise in theories with large extra-dimensions \cite{extra},
as well as in some grand-unified theories like $E_6$. The 
presence of vector-like quarks leads to a small deviation of
$3 \times 3$ unitarity of $V_{CKM}$ which in turn leads to Z-mediated 
new contributions to $B_{d}^{0}-\overline{B}_{d}^{0}$ and 
$B_{s}^{0}-\overline{B}_{s}^{0}$ mixings. In the minimal 
Supersymmetric Standard Model (MSSM) the size of SUSY
contributions to $B_{d}^{0}-\overline{B}_{d}^{0}$  and 
$B_{s}^{0}-\overline{B}_{s}^{0}$ mixing crucially depends
on the choice of soft-breaking terms, but there 
is a wide range of the parameter space where SUSY
contributions can be significant. Recently, it has been
pointed out \cite{Chang:2002mq} that in the context of SUSY SO(10),
there is an interesting connection between the observed large 
mixing in atmospheric neutrinos and the size of the SUSY
contribution to  $B_{s}^{0}-\overline{B}_{s}^{0}$ mixing,
which is expected to be large in this class of models.

The standard way of testing the compatibility of the SM with
the existing data consists of adopting the Wolfenstein parametrization and
plotting in the $\rho $, $\eta $ plane the constraints derived from various
experimental inputs, like the value of $\varepsilon _{K}$, the size of 
$\left| V_{ub}\right| /\left| V_{cb}\right| $,
the value of $a_ {J/\psi K_s}$, as well as the strength of 
$B_{d}^{0}-\overline{B}_{d}^{0}$ 
and $B_{s}^{0}-\overline{B}_{s}^{0}$ mixings. The challenge 
for the SM is then to find a region in 
the $\rho $, $\eta $ plane where all the constraints are simultaneously 
satisfied.

In this paper, we will choose a complete set of rephasing invariant
phases and use $3 \times 3$ unitarity of $V_{CKM}$ to derive a set
of exact relations written in terms of measurable quantities,
namely moduli of $V_{CKM}$  and arguments of rephasing invariant
quartets. We will point out that these exact relations can play 
an important r\^ ole in complementing the standard analysis
in the  $\rho $, $\eta $ plane. Since all relations are exact and
written in terms of measurable quantities, they are particularly
suited to perform precise tests of the SM. Apart from providing
stringent tests of the SM, these exact relations can be useful in finding
the nature of NP. Let us assume that the presence of NP is established
by the impossibility of finding a region in the  $\rho $, $\eta $ plane
where all experimental data can be fitted, within the SM. This
will not indicate which one of the measurements of the angles or the
sides of the unitarity triangle were affected by the presence of
NP. The knowledge of which of the exact relations are
violated by the data may be useful for discovering the source of NP.
By assuming a certain level of precision in future data arising from
the various B-factories, we estimate the power of these exact
relations in either putting bounds on the the strength 
of NP or in revealing its presence.

\section{Choice of Rephasing Invariant phases}

By using the freedom to rephase quark fields, one can readily show that in
the $3\times 3$ sector of a CKM matrix of arbitrary size there are only four
independent rephasing invariant phases. 
Note that this result is completely
general, it does not depend on the number of 
generations and holds true even if
the CKM matrix is not unitary \cite{Branco:1999fs}. 
The number four, is obtained by observing
that in the $3\times 3$ sector of a CKM matrix there are, obviously, nine
phases, and five of them can be removed by rephasing quark fields. We will
choose the following rephasing invariant phases:
\begin{equation}
\begin{array}{c}
\gamma \equiv \func{arg}(-V_{ud}V_{cb}V_{ub}^{\ast }V_{cd}^{\ast })=\func{arg}\left( -\frac{V_{ud}V_{ub}^{\ast }}{V_{cd}V_{cb}^{\ast }}\right)  \\ 
\beta \equiv \func{arg}(-V_{cd}V_{tb}V_{cb}^{\ast }V_{td}^{\ast })=\func{arg}
\left( -\frac{V_{cd}V_{cb}^{\ast }}{V_{td}V_{tb}^{\ast }}\right)  \\ 
\chi \equiv \func{arg}(-V_{cb}V_{ts}V_{cs}^{\ast }V_{tb}^{\ast })=\func{arg}\left( -\frac{V_{cb}V_{cs}^{\ast }}{V_{tb}V_{ts}^{\ast }}\right)  \\ 
\chi ^{\prime }\equiv \func{arg}(-V_{us}V_{cd}V_{ud}^{\ast }V_{cs}^{\ast })=\func{arg}\left( -\frac{V_{us}V_{ud}^{\ast }}{V_{cs}V_{cd}^{\ast }}\right) 
\end{array}
\label{fases}
\end{equation}
Notice that $\chi $ is frequently denoted in 
the literature as $\beta _{s}$ and $\chi ^{\prime }$ as $\beta _{K}$. 
It is important to stress that since these phases are rephasing invariant
quantities, they correspond to physical observables. If one assumes $3\times
3$ unitarity of $V_{CKM}$, it has been shown that one can reconstruct the
full CKM matrix, using the above four rephasing invariant phases as input
\cite{Aleksan:1994if}.
At this stage, the following comment is in order. Having in mind that it has
also been shown \cite{Branco:1987mj} that one can 
reconstruct the full CKM matrix from four
independent moduli, one may test the SM by comparing the unitarity triangles
obtained from the measurement of the four independent phases 
$\gamma $, $\beta $, $\chi $, $\chi ^{\prime }$ with the unitary 
triangles obtained from
the knowledge of four independent moduli, which can be chosen to be 
$\left|V_{us}\right| $, $\left| V_{cb}\right| $, 
$\left| V_{ub}\right| $, $\left|V_{td}\right| $. Although such a test 
is possible in theory, its practical
interest is limited by the fact that at least one of the phases, 
namely $\chi ^{\prime }$, is probably too small to be measurable 
through B decays and furthermore the
extraction of $\left| V_{td}\right| $ and $\left| V_{ub}\right| $ from
experiment is plagued by hadronic uncertainties. Obviously, 
within the framework of the SM,
$\chi ^{\prime }$ can be obtained from the
experimental value of $\varepsilon _{K}$, but its extraction
also suffers from hadronic uncertainties.  

In order to fix the invariant phases entering in $B^{0}$ CP asymmetries, it
is useful to adopt the following phase convention, \cite{Branco:1999fs}:
\begin{equation}
\func{arg}(V)=\left( 
\begin{array}{lll}
0 & {\chi ^{\prime }} & -\gamma  \\ 
\pi  & 0 & 0 \\ 
-\beta  & \pi +\chi  & 0
\end{array}
\right) \label{fasesconv}
\end{equation}
Through the measurement of CP asymmetries, one can obtain the phases of the
rephasing invariant quantities:
\begin{equation}
\lambda _{f}^{\left( q\right) }=\left( \frac{q_{B_{q}}}{p_{B_{q}}}\right)
\left( \frac{A\left( \overline{B}_{q}^{0}\rightarrow f\right) }{A\left(
B_{q}^{0}\rightarrow f\right) }\right) 
\text{ ; }\lambda _{\overline{f}}^{\left( q\right) }=
\left( \frac{q_{B_{q}}}{p_{B_{q}}}\right) \left( 
\frac{A\left( \overline{B}_{q}^{0}\rightarrow \overline{f}\right) }
{A\left(B_{q}^{0}\rightarrow \overline{f}\right) }\right) \label{lambdas}
\end{equation}
The first factor in $\lambda _{f}^{\left( q\right) }$ is due to mixing and
its phase equals $(-2\beta )$ and $2\chi $ for $B_{d}$ and $B_{s}$,
respectively.
Later on, we will assume that NP can give contributions to mixing. Then, it
is useful to parametrize these NP contributions in the following way 
\begin{eqnarray}
M_{12}^{\left( q\right) } &=&\left( M_{12}^{\left( q\right) }\right)
^{SM}r_{q}^{2}e^{-2i\phi _{q}}\Rightarrow \Delta M_{B_{q}}=\left( \Delta
M_{B_{q}}\right) ^{SM}r_{q}^{2} \label{Munodos}\\
&&  \notag \\
\frac{q_{B_{q}}}{p_{B_{q}}} &=&\exp \left( i\arg \left( M_{12}^{\left(
q\right) }\right) ^{\ast }\right) =\left( \frac{q_{B_{q}}}{p_{B_{q}}}\right)
^{SM}e^{2i\phi _{q}}\label{mixing}
\end{eqnarray}
In the presence of NP, the phases from mixing become 
$2(-\beta +\phi_{d})\equiv - 2\overline{\beta }$ and 
$2(\chi +\phi _{s})\equiv 2\overline{\chi }$ for 
$B_{d}$ and $B_{s}$ decays, respectively. It is clear that $r_{q}\neq 1$ 
and/or $\phi _{q}\neq 0$ would signal the presence of NP. How
could one measure $\beta $ without contamination of NP? One possibility
would be to measure $\beta $ through the t-quark contribution to the $%
b\longrightarrow d$ penguin, since this contribution is proportional to $%
V_{tb}V_{td}^{\ast }$, which in our phase convention is $\beta $.
Unfortunately, this is not possible \cite{London:1999iv} without theoretical 
input about hadronic
parameters which involve large uncertainties. This is essentially due to the
fact that the $b\longrightarrow d$ penguin receives significant u and c
quark contributions. Concerning $\chi $, in general, the existence of NP
contributions to mixing in the $B_{s}$ system would no longer lead to a zero
asymmetry, as predicted in the framework of the SM, in the $B_{s}$ decay
dominated by the $b\longrightarrow s$ penguin such as $B_{s}\rightarrow \phi
\eta ^{\prime }$. In this case, a deviation from zero in this asymmetry would
be a clear indication of NP also implying that the CP asymmetry in the
channel $B_{s}\rightarrow D_{s}\overline{D_{s}}$ will measure directly 
a $\overline{\chi }$ which may differ from $\chi $. 
The difficulty of separating $\beta $ from a possible NP contribution $\phi
_{d}$ in ${B^{0}}_{d}$ decays like ${B^{0}}_{d}\rightarrow J/\Psi K_{s}$,
renders specially important the measurement of $\gamma $, which does not
suffer from contamination of NP in the mixing. Note that $\gamma $ can be
either directly measured \cite{gamma} 
or obtained through the knowledge of the
asymmetries $a_{J/\Psi K_{s}}=\func{Im}\left( 
\lambda _{J/\Psi K_{s}}^{\left( d\right) }\right) $, 
$a_{\pi ^{+}\pi ^{-}}=\func{Im}\left(
\lambda _{\pi ^{+}\pi ^{-}}^{\left( q\right) }\right) $. 
Indeed the phase $\phi _{d}$ cancels in the sum 
$\overline{\alpha }+\overline{\beta }=\left(
\pi -\gamma -\beta +\phi _{d}\right) +\left( \beta -\phi _{d}\right) $ and
one has:
\begin{equation}
\gamma =\pi -\frac{1}{2}\left[ \arcsin a_{J/\Psi K_{s}}+\arcsin a_{\pi
^{+}\pi ^{-}}\right]\label{arcsin} 
\end{equation}
Note that we are using $a_{\pi^{+}\pi ^{-}} = \sin (2 \overline{\alpha})$
that can be extracted from the experimental asymmetry through
various different approaches \cite{app}.
Once $\gamma $ is known, $\beta $ can be readily obtained, using unitarity
and the knowledge of $\left| V_{ub}\right| $, $\left| V_{us}\right| $, $\left| V_{cb}\right| $. The knowledge of $\beta $, together with $a_{J/\Psi
K_{s}}$ leads then to the determination of $\phi _{d}$. Of course, this
evaluation of $\phi _{d}$ will be restricted by the precision on $\left|
V_{ub}\right| $, since $\left| V_{us}\right| $, $\left| V_{cb}\right| $ are
extracted from experiment with good accuracy. Similar considerations apply to
the extraction of $r_{d}$, $r_{s}$ or $r_{d}/r_{s}$ from $\Delta M_{B_{d}}$
and $\Delta M_{B_{s}}$ where $\left| V_{td}^{\ast }V_{tb}\right| $, 
$\left| V_{ts}^{\ast }V_{tb}\right| $ or its ratio, 
have to be reconstructed previously using unitarity.

In Table 1 we summarize the phases that can be measured 
from CP asymmetries both in the KM scheme or in the case of 
New Physics in the mixing. For example, $\overline{\beta } $
is the phase measured through $a_{J/\Psi K_{s}}$, i.e., 
$\overline{\beta } = (1/2) \arcsin a_{J/\Psi K_{s}}$.

\begin{table}[tbp] \centering%
\centering
\begin{tabular}{|c|c|c|c|c|c|}
\hline
K-M & $\beta $ & $\gamma $ & $\chi $ & $\chi ^{\prime }$ & $\alpha =\pi
-\beta -\gamma $ \\ \hline
N-P & $\overline{\beta }=\beta -\phi _{d}$ & $\gamma $ & $\overline{\chi }%
=\chi +\phi _{s}$ & $\chi ^{\prime }$ & $\overline{\alpha }=\alpha +\phi _{d}
$ \\ \hline
\end{tabular}
\caption{Phases that can be measured in different scenarios}
\label{tabla1}
\end{table}

\section{Precision Tests of the SM and search for New Physics}

In this section, we derive a complete set of exact relations
involving moduli of $V_{CKM}$ and the four rephasing invariant
phases of Eq.(\ref{fases}) adopting the phase
convention of Eq.(\ref{fasesconv}).
From the six unitarity relations corresponding 
to orthogonality of different rows and of different columns 
of $V_{CKM}$ one obtains: 
\begin{eqnarray}
(uc) &&\sin \chi ^{\prime }=\frac{\left| V_{ub}\right| \left| V_{cb}\right| 
}{\left| V_{us}\right| \left| V_{cs}\right| }\sin \gamma  \label{rel3} \\
&&  \notag \\
(ut) &&\left| V_{ud}\right| \left| V_{td}\right| \sin \beta -\left|
V_{us}\right| \left| V_{ts}\right| \sin (\chi ^{\prime }-\chi )-\left|
V_{ub}\right| \left| V_{tb}\right| \sin \gamma =0  \label{rel4} \\
&&  \notag \\
(ct) &&\sin \chi =\frac{\left| V_{td}\right| \left| V_{cd}\right| }{\left|
V_{ts}\right| \left| V_{cs}\right| }\sin \beta  \label{rel5} \\
&&  \notag \\
(db) &&\frac{\left| V_{ub}\right| }{\left| V_{td}\right| }=\frac{\sin \beta 
}{\sin \gamma }\frac{\left| V_{tb}\right| }{\left| V_{ud}\right| }
\label{rel6} 
%&&  \notag \\
\end{eqnarray}
\begin{eqnarray}
(ds) &&\sin \chi ^{\prime }=\frac{\left| V_{td}\right| \left| V_{ts}\right| 
}{\left| V_{ud}\right| \left| V_{us}\right| }\sin (\beta +\chi )
\label{rel7} \\
&&  \notag \\
(sb) &&\frac{\sin \chi }{\sin (\gamma +\chi ^{\prime })}=\frac{\left|
V_{us}\right| \left| V_{ub}\right| }{\left| V_{ts}\right| \left|
V_{tb}\right| }  \label{rel8}
\end{eqnarray}
where, in parenthesis, we have indicated the corresponding 
rows and columns. There are additional relations which 
can be readily obtained either by orthogonality or by applying the
law of sines to the corresponding unitarity triangles, such as: 
\begin{eqnarray}
(db) &&\left| V_{ub}\right| =\frac{\left| V_{cd}\right| \left| V_{cb}\right| 
}{\left| V_{ud}\right| }\frac{\sin \beta }{\sin (\gamma +\beta )}
\label{rel9} \\
&&  \notag \\
(db) &&\left| V_{td}\right| =\frac{\left| V_{cd}\right| \left| V_{cb}\right| 
}{\left| V_{tb}\right| }\frac{\sin \gamma }{\sin (\gamma +\beta )}
\label{rel10} \\
&&  \notag \\
(sb) &&\sin \chi =\frac{\left| V_{us}\right| \left| V_{ub}\right| }{\left|
V_{cs}\right| \left| V_{cb}\right| }\sin (-\chi +\chi ^{\prime }+\gamma )
\label{rel11}
\end{eqnarray}
Furthermore, by dividing Eq.(\ref{rel10}) by $\left| V_{ts}\right| $ and
using normalization of rows and columns one obtains 
\begin{equation}
r=\frac{\sin \gamma }{\sin (\gamma +\beta )}\frac{\left| V_{cd}\right| }
{\left| V_{tb}\right| }\left[ 1+r^{2}-r^{2}\frac{\sin ^{2}\beta }{\sin
^{2}\gamma }\frac{\left| V_{tb}\right| ^{2}}{\left| V_{ud}\right| ^{2}}
\right] ^{\frac{1}{2}}  \label{rel13}
\end{equation}
where $r\equiv \left| V_{td}\right| /\left| V_{ts}\right| $. Using Eqs.(\ref
{rel4}) and (\ref{rel6}), together with normalization conditions, one
obtains: 
\begin{equation}
\sin (\chi -\chi ^{\prime })=r\sin \beta \frac{\left| V_{us}\right| }{\left|
V_{ud}\right| }\left[ 1-\frac{\left| V_{cb}\right| ^{2}}{\left|
V_{us}\right| ^{2}}\right]  \label{rel14}
\end{equation}
Another interesting relation is obtained by combining Eq.(\ref{rel8}) with
Eq.(\ref{rel9}), leading to: 
\begin{equation}
\sin \chi =\frac{\left| V_{us}\right| \left| V_{cd}\right| \left|
V_{cb}\right| }{\left| V_{ts}\right| \left| V_{tb}\right| \left|
V_{ud}\right| }\frac{\sin \beta \sin (\gamma +\chi ^{\prime })}{\sin (\gamma
+\beta )}  \label{rel15}
\end{equation}
Since the above formulae have the potential of providing precise tests of
the SM, we have opted for writing exact relations. However, it is obvious
that given the experimental knowledge on the size of the various moduli of
the CKM matrix elements, some of the above relations can be, to an excellent
approximation, substituted by simpler ones. For example, Eq.(\ref{rel15}) is
the exact version of the Aleksan-London-Kayser relation \cite{Aleksan:1994if}, 
the importance of which has been 
emphasized by Silva and Wolfenstein \cite{Silva:1996ih} : 
\begin{equation}
\sin \chi \simeq \frac{\left| V_{us}\right| ^{2}}{\left| V_{ud}\right| ^{2}}
\frac{\sin \beta \sin \gamma }{\sin (\gamma +\beta )}  \label{rel16}
\end{equation}
Similarly Eq.(\ref{rel5}) can be well approximated by:
\begin{equation}
\sin \chi \simeq r \frac {\left| V_{us}\right| } {\left| V_{ud}\right| }
\sin \beta  \label{rel17}
\end{equation}
and Eqs.(\ref{rel10}) and (\ref{rel8}) lead, respectively to: 
\begin{eqnarray}
r &\simeq &\left| V_{us}\right| \frac{\sin \gamma }{\sin (\gamma +\beta )}
\label{rel18} \\
\sin \chi &\simeq &\frac{\left| V_{us}\right| \left| V_{ub}\right| }{\left|
V_{cb}\right| }\sin \gamma  \label{rel19}
\end{eqnarray}
At this stage, the following comments are in order:
\begin{itemize}
\item[(i)]  Eq.(\ref{rel3}) would provide an excellent test of the SM if the
phase $\chi ^\prime $ could be measured without hadronic uncertainties.
As previously mentioned, $\chi ^\prime $ can be obtained from the
knowledge of $\varepsilon _{K}$, but its extraction suffers from
hadronic uncertainties in  ${f_K}^2 B_K$.  
Note that all quantities in Eq.(\ref
{rel3}) are immune to the presence of New Physics in 
$B_{d}^{0}-\overline{B}_{d}^{0}$ and 
$B_{s}^{0}-\overline{B}_{s}^{0}$ mixings.
\item[(ii)]  Eq.(\ref{rel5}) and its approximate form Eq.(\ref{rel17}) would
provide an excellent test of the SM, once $\chi $, $r$ and $\beta $ are
measured. Note that the theoretical errors in extracting $r\equiv \left|
V_{td}\right| /\left| V_{ts}\right| $ from $B_{d}^{0}-\overline{B}_{d}^{0}$
and $B_{s}^{0}-\overline{B}_{s}^{0}$ mixings are much smaller than those
present in the extraction of $\left| V_{td}\right| $, $\left| V_{ts}\right| $.
\item[(iii)]  Eq.(\ref{rel19}) has the important feature of only involving
quantities which are not sensitive to the possible presence of New Physics
in $B_{d}^{0}-\overline{B}_{d}^{0}$ mixing. It has, of course, the
disadvantage of requiring the knowledge of $\left| V_{ub}\right| $ with
significant precision, in order to be a precise test of the SM.
\item[(iv)]  Eq.(\ref{rel13}) in an exact way and Eq.(\ref{rel18}) in an
excellent approximation, give $r$ in terms of $\gamma $ and $\beta $. This
relation will provide an important test of the SM once $r$, $\gamma $ and $
\beta $ are measured. Note that in the SM, one knows that $r$ is of order 
$\left| V_{us}\right| $, the importance of Eq.(\ref{rel18}) is that it
provides the constant of proportionality.
\end{itemize}
In the context of the SM the above formulae can 
also be very useful for a precise
determination of $V_{CKM}$ from input data: for example, if $\beta $ and 
$\gamma $ are measured with sufficient accuracy, one can use 
Eqs.(\ref{rel9}), (\ref{rel10}) to determine $\left| V_{ub}\right| $, 
$\left| V_{td}\right| $. 
One can thus reconstruct the full CKM matrix, using $\left| V_{us}\right| 
$, $\left| V_{cb}\right| $, $\beta $ and $\gamma $ as input parameters.
Furthermore we can also predict the SM value for $\sin 2\chi $ and $\sin
\chi ^{\prime }$.

From Eq.(\ref{rel5}) we can write 
\begin{equation}
\begin{array}{ccc}
\sin \chi \simeq R_{t}\left| V_{us}\right| ^{2}\sin \beta  & ; & R_{t}=\frac{
\left| V_{td}\right| \left| V_{tb}\right| }{\left| V_{cd}\right| \left|
V_{cb}\right| }
\end{array}
\label{sinkappa}
\end{equation}
leading to 
\begin{equation}
0.030\leq \sin 2\chi \equiv \sin 2\beta _{s}\leq 0.045  \label{sinkappan}
\end{equation}
using the present knowledge of the parameters \cite{bbb} ,
$R_{t}=0.84\pm 0.08$ , $\left| V_{us}\right| =0.221\pm 0.002$ 
and $\beta =\left( 26.9\pm 5.0\right) 
%TCIMACRO{\UNICODE[m]{0xba}}%
%BeginExpansion
{{}^o}%
%EndExpansion
$ . The SM interval for $\sin \chi ^{\prime }$ can be obtained from either 
Eq.(\ref{rel3}) or Eq.(\ref{rel7}), its central 
value is predicted to be much
smaller than the one of $\sin \chi $. Although $\gamma $ has not yet been
measured, the allowed interval in the SM is \cite{bbb}
\begin{equation}
\gamma =\left( 55.4\pm 11.9\right) 
{{}^o}
\label{gamma}
\end{equation}
Using Eq.(\ref{gamma})together with the present values \cite{bbb}
$\left| V_{cb}\right| =0.0417\pm 0.0010$,
$\left| V_{ub}\right| =\left( 4.05\pm 0.42\right) \times 10^{-3}$, we
obtain 
\begin{equation}
5.4\times 10^{-4}\leq \sin \chi ^{\prime }\leq 8.2\times 10^{-4}
\label{kappap}
\end{equation}

So far our discussion has been done in the framework of the SM. 
Let us now consider Physics Beyond the SM.There is a large class 
of extensions of the SM in which the CKM matrix
remains unitary and therefore the unitarity relations previously written in
this section are still valid. An important example is , of course , the case
of supersymmetric extensions of the SM. We will assume that New Physics only
contributes to one loop processes such as $B_{d}^{0}-\overline{B}_{d}^{0}$
and $B_{s}^{0}-\overline{B}_{s}^{0}$ mixing, so that the measurement of 
$\Delta M_{B_{d}}$ and $\Delta M_{B_{s}}$ will not allow us to extract
directly the values of $\left| V_{td}V_{tb}\right| $ and 
$\left| V_{ts}V_{tb}\right| $. Furthermore, CP asymmetries 
will no longer measure
the same angles as in the framework of the SM. As a result, a naive
extraction of the values of the sides and angles of the unitarity triangles
from input experimental data ( i.e. assuming the validity of the SM ) will
not lead to the correct values 
\footnote{It is interesting to note that the naive extraction 
of the same parameter through different processes affected in a different 
way by NP may not lead to a unique value, thus signalling NP}. 
This implies that although the
unitarity relations continue to be valid, the presence of NP will simulate
violations of the above relations.
In the presence of NP one may wonder how large the deviations from
the SM should be in order to be possible to detect them using these
unitarity relations. This NP scenario is 
parametrized by Eq.(\ref{Munodos}) and
the phases entering in each process are defined in Table \ref{tabla1}. The
magnitudes that will signal NP are $\phi _{d},\phi _{s}\neq 0$ and/or  
$r_{d},r_{s}\neq 1$. 

\vspace{0.5cm}

{\bf Extraction of $\phi_{d}$}

From Eq.(\ref{rel9}), we see that this unitarity relation
can only be affected by the presence of $\phi _{d}$, therefore this
equation allows for a clean extraction of $\phi _{d}$. 
By writing Eq.(\ref{rel9}) in terms of $\overline{\beta }$ 
and $\phi _{d}$ (note that $Im\left( \lambda _{J/\Psi K_{s}}^{(d)}\right)
 =\sin \left( 2\overline{\beta } \right) $) we get 
\begin{equation}
\tan \left( \phi _{d}\right) =\frac{R_{u}\sin \left( \gamma +\overline{\beta 
}\right) -\sin \left( \overline{\beta }\right) }{\cos \left( \overline{\beta 
}\right) -R_{u}\cos \left( \gamma +\overline{\beta }\right) }  \label{phid}
\end{equation}
with 
\begin{equation}
R_{u}=\frac{\left| V_{ud}\right| \left| V_{ub}\right| }{\left| V_{cd}\right|
\left| V_{cb}\right| }  \label{Ru}
\end{equation}

From Eq.(\ref{phid}), we can find out the bounds that can be reached for 
$\phi _{d}$, once we have a direct measurement of $\gamma $. 
To illustrate the usefulness of Eqs.(\ref{rel9}) and 
(\ref{phid}), we will give examples of different sets of assumed 
data which hopefully will be available in the near future. We will 
assume that apart from $\left| V_{us}\right|$, $\left| V_{cb}\right|$,
$\left| V_{ub}\right|$  and $\overline{\beta }$ also $\gamma $ 
has been measured. We will consider three different cases. In
the first case (Example 1), we assume data with relatively large
errors for $\overline{\beta }$, $\gamma $ and $\left| V_{ub}\right|$,
leading to a value $\phi _{d}$ with relatively large errors and consistent
with zero. In the second case (Example 2), we assume data with smaller
errors, leading to a more precise determination of $\phi _{d}$, but
still consistent with zero. This is, of course, the scenario where
no NP is discovered and strict bounds are set on the strength of NP.
Finally in the third case (Example 3) we consider again data with 
small errors but leading to a value of $\phi _{d}$ which is not
consistent with zero. This is the most optimistic scenario, where NP
is discovered. For each set of input data we depict the
corresponding $\phi _{d}$ distribution, generated by means of a toy
Monte Carlo, with the assumption of Gaussian errors.

EXAMPLE 1: In our first example we use as input values
 \begin{equation}
\begin{array}{ccc}
\left| V_{us}\right| =0.221\pm 0.002 & \left| V_{cb}\right| =0.0417\pm 0.0010
& \left| V_{ub}\right| =\left( 4.05\pm 0.42\right) \times 10^{-3} \\ 
{\overline \beta} =\left( 26.9\pm 5.0\right)
{{}^o}
& \gamma =\left( 55.4\pm 11.9\right)
{{}^o}
& 
\end{array}
\label{acval}
\end{equation}
\begin{figure}[t]
\begin{center}
\epsfig{file=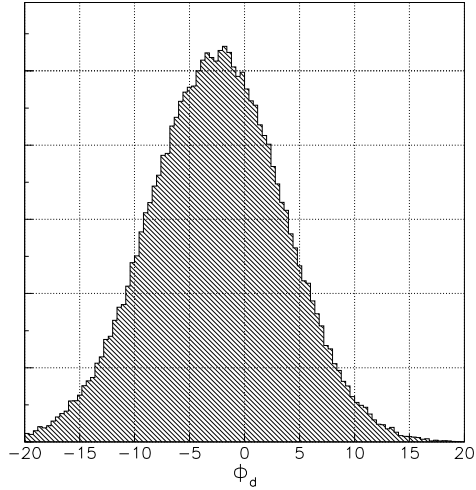,width=2.8781in}
\end{center}
\caption{The $\protect\phi _{d}$
distribution in degrees corresponding to EXAMPLE 1, consistent
with $\protect\phi _{d}=0$.}
\label{figura1}
\end{figure}
The corresponding distribution is given by Fig.\ref{figura1}. We conclude
that if we assume the present experimental numbers and a poor 
determination of $\gamma $ ($\sim 20\%$ error), one obtains 
$\phi_{d}=\left( -2.6\pm 6.0\right){{}^o}$, consistent with zero.

\begin{figure}[t]
\begin{center}
\epsfig{file=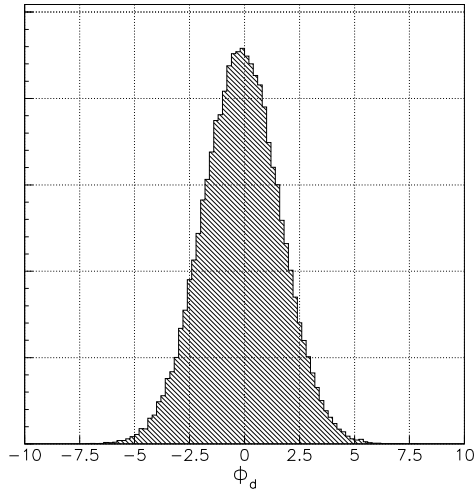,width=2.8781in}
\end{center}
\caption{The $\protect\phi _{d}$
distribution in degrees corresponding to EXAMPLE 2, consistent
with $\protect\phi _{d}=0$.}
\label{figura2}
\end{figure}

EXAMPLE 2: In our second example we assume a knowledge of 
${\overline \beta}$, $\gamma $ and $\left| V_{ub}\right|$ at
the level of  $(1\%)$,  $(10\%)$ and  $(5\%)$, respectively.
This level of precision is expected in future B factories. In this case
we use as input values 
\begin{equation}
\begin{array}{ccc}
\left| V_{us}\right| =0.221\pm 0.002 & \left| V_{cb}\right| =0.0417\pm 0.0010
& \left| V_{ub}\right| =\left( 4.05\pm 0.21\right) \times 10^{-3} \\ 
{\overline \beta} =\left( 25.1\pm 0.25\right) 
{{}^o}
& \gamma =\left( 56.6\pm 5.6\right) 
{{}^o}
& 
\end{array}
\label{fuval}
\end{equation}
and the corresponding distribution is given by Fig.\ref{figura2}.
This example leads to 
$\phi_{d}=\left( -0.1\pm 1.7\right){{}^o}$ which is, once again,
consistent with zero. We conclude that discovering NP contributing
to the phase of $B_{d}^{0}-\overline{B}_{d}^{0}$ mixing 
will be possible for values of $\phi_{d}$ of a few degrees.

Note that for  $\gamma +\overline{\beta }$ close to $90{{}^o}$, 
as is the case in our two previous examples, the dependence of
$\phi_{d}$ on the precise value of  $\gamma +\overline{\beta }$ 
is rather weak so that, in these cases, it will be the precision
of $\left| V_{ub}\right| $ that will control the final bound on
$\phi_{d}$. Obviously, this will no longer be true if 
$\gamma +\overline{\beta }$  differs significantly from
$90{{}^o}$.

\begin{figure}[h]
\begin{center}
\epsfig{file=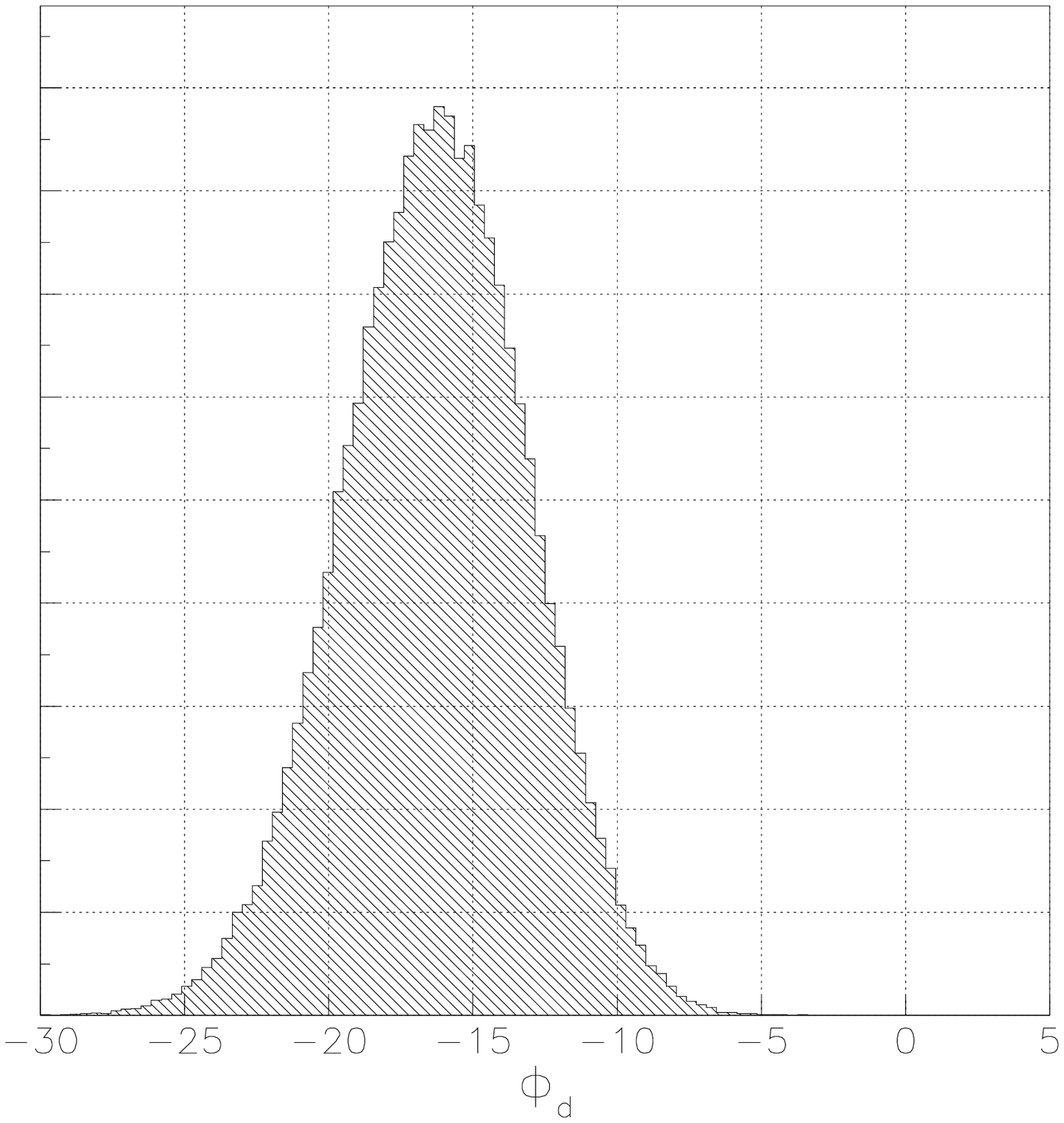,width=2.8781in}
\end{center}
\caption{The $\protect\phi _{d}$
distribution in degrees corresponding to EXAMPLE 3, pointing
clearly to NP.}
\label{fidnpcaso1}
\end{figure}

EXAMPLE 3: So far we have only considered examples leading to 
$\phi_{d}$ consistent with zero. Let us now consider an example
that would lead to a sizeable $\phi_{d}$. If we replace the values of
$\overline{\beta }$ and $\gamma$ in Eq.(\ref{fuval}) by:
\begin{equation}
\overline \beta=\left(30.0\pm 0.3\right){{}^o} \ \ \  
\gamma=\left(20\pm 5\right){{}^o}
\end{equation}
the resulting $\phi_{d}$ distribution is presented in 
Fig.\ref{fidnpcaso1} corresponding
to $\phi_{d}=\left( -16.3 \pm 3.2\right){{}^o}$
In this case one would have a clear indication 
of NP in the phase of $B_{d}^{0}-\overline{B}_{d}^{0}$ mixing.
Note that for this choice of $\gamma$ the value of 
$\varepsilon _{K}$ would not be saturated by the SM
contribution. Therefore in this example one would conclude 
that NP also contributes to 
$\varepsilon _{K}$.

\vspace{0.5 cm}

{\bf Extraction of $\phi _s$}

\begin{figure}[h]
\begin{center}
\epsfig{file=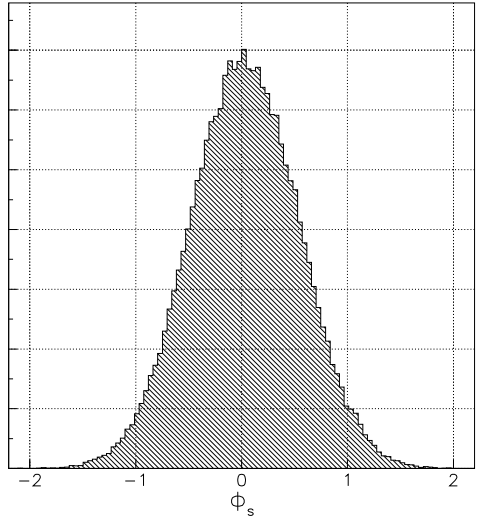,width=3.0493in}
\end{center}
\caption{The $\protect\phi _{s}$
distribution in degrees for the assumed values of $\protect\gamma =\left(
56.6\pm 5.6\right)^o$ and $\overline{\protect\chi }=\left(
1.06\pm 0.50\right)^o$. Other inputs are the values 
given in Eq.(\ref{acval}).}
\label{figura3}
\end{figure}

New Physics in the phase of  $B_{s}^{0}-\overline{B}_{s}^{0}$ mixing 
would be seen through Eq.(\ref{rel11}) where all 
variables except $\chi$ (and eventually $\chi ^{\prime }$
which is too small to play any r\^ ole) can be extract
from experiment independently of NP. This equation allows for
a clean extraction of $\phi _{s}$ \footnote {We are 
assuming that we are not trying to disentangle $\chi $ from 
$\phi _{s}$ with the measurement of an asymmetry directly sensitive 
just to $\phi _{s}$} through the measurement of an asymmetry
sensitive to $\overline{\chi }$. Neglecting $\chi ^{\prime }$
and rewriting Eq.(\ref{rel11}) in terms of $\overline{\chi }$
and $\phi _{s}$ we get
\begin{equation}
\tan \left( \phi _{s}\right) =\frac{\sin \overline{\chi }-C\sin \left(
\gamma -\overline{\chi }\right) }{C\cos \left( \gamma -\overline{\chi }
\right) +\cos \overline{\chi }}  \label{phis}
\end{equation}
with
\begin{equation}
C=\frac{\left| V_{us}\right| \left| V_{ub}\right| }{\left| V_{cs}\right|
\left| V_{cb}\right| }  \label{Ce}
\end{equation}
Applying the previous procedure to Eq.(\ref{phis}) with the 
choice of assumed data given by $\left| V_{us}\right| $,
$\left| V_{cb}\right| $ and $\left| V_{ub}\right| $ in 
Eq.(\ref{acval}) together with
\begin{equation}
\gamma =\left(56.6\pm 5.6\right)^o \ \ \ 
\overline{\chi }=\left( 1.06\pm 0.50\right)^o
\end{equation}
we are led to  $\phi _{s}=\left( 0.03\pm 0.51\right){{}^o}$,
the corresponding Gaussian distribution is given by Fig.\ref{figura3}.
In this example we assumed $\gamma $ and $\overline{\chi }$ to be 
known to a precision of  $20\%$ and $50\%$ level respectively.
We conclude that it will be possible to discover NP
contributing to the phase of  $B_{s}^{0}- \overline{B}_{s}^{0}$ 
mixing for values of $\phi_{s}$ larger than about  $1.5{{}^o}$.
%\FRAME{ftbpFU}{3.0493in}{3.0493in}{0pt}{\Qcb{The $\protect\phi _{s}$
%distribution in degrees for the assumed values of $\protect\gamma =\left(
%55.4\pm 11.9\right)
%TCIMACRO{\UNICODE[m]{0xba}}%
%BeginExpansion
%{{}^o}%
%EndExpansion
%$ and $\overline{\protect\chi }=\left( 1.06\pm 0.20\right)
%TCIMACRO{\UNICODE[m]{0xba}}%
%BeginExpansion
%{{}^o}%
%EndExpansion
%$. Other inputs are taken to its actual values.}}{\Qlb{figura3}}{fis-0.eps}{%
%\special{language "Scientific Word";type "GRAPHIC";maintain-aspect-ratio
%TRUE;display "USEDEF";valid_file "F";width 3.0493in;height 3.0493in;depth
%0pt;original-width 7.8793in;original-height 7.8793in;cropleft "0";croptop
%"1";cropright "1";cropbottom "0";filename 'global/fis-0.eps';file-properties
%"XNPEU";}}

\vspace{0.5 cm}

{\bf Extraction of $r_d$}

Let us consider the possibility of having $r_d$ and/or $r_s$
different from 1, which would also be a sign of NP. 
From Eq.(\ref{rel6}) we get 
\begin{equation}
\left| V_{td}\right| \left| V_{tb}\right| =\frac{\sin \gamma }{\sin \beta }
\left| V_{ud}\right| \left| V_{ub}\right|   \label{VtdVtb}
\end{equation}
With the measurement of $\overline{\beta}$, $\gamma$ and $R_u$ the value of 
$\phi_d$ can be determined from Eq.(\ref{phid}). As a result one can 
obtain $\beta = \overline{\beta} -\phi_d $ and use Eq.(\ref{VtdVtb})
to determine  $\left| V_{td}\right| \left| V_{tb}\right| $
This  determination of $\left| V_{td}\right| \left| V_{tb}\right| $
should be compared to the value of  $\left|
V_{td}\right| \left| V_{tb}\right| r_d $ extracted from the
experimental value of $\Delta M_{B_{d}}$, in order to obtain $r_d$.

From B-factories, CLEO and Tevatron one can expect, as mentioned before,
significant improvements in the level of precision to which 
$\overline \beta$ and  $\left| V_{ub}\right| $ 
are determined (to $1\%$ and $5\%$
respectively) together with a direct measurement of  $\gamma $
with about $10\%$ precision.
Therefore, we can expect to have a determination of the left hand
side (lhs) of Eq.(\ref{VtdVtb}) at the $7\%$ level if the central
value of $\gamma$ does not deviate significantly from the 
presently allowed region within the SM.
The largest uncertainty comes from the extraction of 
$\left| V_{td}\right| \left| V_{tb}\right| r_{d}$ from 
$\Delta M_{B_{d}}$,
which is theoretically limited by the presence of the non-perturbative
parameter $f_{B_{d}}B_{B_{d}}^{1/2}=\left( 230\pm 34\right) MeV$.
With the present theoretical knowledge one cannot
expect a determination of $r_{d}$ with a precision better than $15-20\%$.
Future improvement depends on more precise lattice results.
Similar arguments would apply to the determination of 
$r_{s}$ from the experimental value of $\Delta M_{B_{s}}$,
which will be available in the near future.
The CP asymmetry in semileptonic B decays also contains
correlated information about $r_d$ and $\phi_d$ 
\cite{Branco:1992uy}, \cite{Laplace:2002ik}.
In fact the present world average 
$A_{SL} =(0.2 \pm 1.4) \times 10^{-2}$  \cite{ASL} allows to put
bounds in the $r_d$, $\phi_d$ plane. In spite of the large
theoretical uncertainties, in future analysis one should 
also include $A_{SL}$ to check the consistency of the extraction
of $\phi_d$ and  $r_d$.

\vspace{0.5 cm}

{\bf Extraction of $r_d /r_s$}

Once the $B_{s}^{0}$
mixing parameter $x_{s}\equiv \Delta M_{B_{s}}\tau _{B_{s}}$
is measured we can directly determine $\left| 
V_{td}\right| r_{d}/\left| V_{ts}\right| r_{s}$ from the 
ratio $\Delta M_{B_{d}}/\Delta M_{B_{s}}$ requiring the 
knowledge of  $\xi \equiv
f_{B_{s}}B_{B_{s}}^{1/2}/f_{B_{d}}B_{B_{d}}^{1/2}=1.16\pm 0.05$.
The parameter $\xi$ is determined from lattice calculations
and suffers from a much smaller uncertainty than each of the
terms in the ratio taken separately.
In this case Eq.(\ref{rel10}) divided by 
$\left| V_{ts}\right| $ will become very useful in the 
determination of $r_d / r_s$. We can rewrite the resulting
right hand side (rhs) of Eq.(\ref{rel10}), to an excellent 
approximation, by replacing $\left| V_{cd}\right| $
by $\left| V_{us}\right| $, $\left| V_{tb}\right| $ by 1
and $\left| V_{ts}\right| $ by 
\begin{equation}
\left| V_{ts} \right| \simeq  \left| V_{cb} \right|
\left\{ 1+\left| V_{us} \right| \left[
\frac{\left| V_{ub} \right| }{\left| V_{cb}\right| } 
\cos \gamma -\frac{1}{2} \left|
V_{us} \right| \right] \right\}
\end{equation}
leading to
\begin{equation}
r\equiv \left| V_{td}\right| /\left| V_{ts}\right| 
= \frac{\left| V_{us} \right| }{ 1+\left| V_{us} \right| \left[
\frac{\left| V_{ub} \right| }{\left| V_{cb}\right| } 
\cos \gamma -\frac{1}{2} \left|
V_{us} \right| \right]} \frac{\sin \gamma}{\sin \left( 
\gamma +\beta \right) } + {\cal O} (\lambda ^5 )
\end{equation}
From this equation together with 
\begin{equation}
\frac{x_{d}}{x_{s}}=r^{2}\left( \frac{r_{d}}{r_{s}}\right) ^{2}\left( \frac{%
M_{B_{d}}\tau _{B_{d}}}{M_{B_{s}}\tau _{B_{s}}}\right) \frac{1}{\xi ^{2}}
\label{xd/xs}
\end{equation}
we obtain
\begin{eqnarray}
\label{frd/rs}
f\left( \phi _{d}\right) \times \left( \frac{r_{d}}{r_{s}}\right)& =&
\left( \xi^{2}\frac{x_{d}M_{B_{s}}\tau _{B_{s}}}
{x_{s}M_{B_{d}}\tau _{B_{d}}}\right)
^{1/2}\left( \frac{\sin \left( \gamma +\overline{\beta }\right) }
{\left| V_{us}\right| \sin \gamma }\right) \times \\ \nonumber 
 & &\times \left( 1+\left| V_{us}\right| \left[
\left| \frac{V_{ub}}{V_{cb}}\right| \cos \gamma -\frac{1}{2}\left|
V_{us}\right| \right] \right)  
\end{eqnarray}
where $f\left( \phi _{d}\right) $ is given by 
\begin{equation}
f\left( \phi _{d}\right) =\left( \cos \phi _{d}\left[ 
1+\frac{\tan \phi _{d}}
{\tan \left( \gamma +\overline{\beta }\right) }\right] \right) ^{-1}
\label{f}
\end{equation}
Equation  (\ref{frd/rs}) is valid to a precision better than  $1\%$.
Obviously, $f\left( \phi _{d}\right) r_{d}/r_{s}=1$ if there is no NP. 
The rhs of Eq.(\ref{frd/rs}) will be evaluated from experimental data
once we have a direct measurement of $\gamma $ and $x_{s}$
thus providing a test of NP in the ratio $r_{d}/r_{s}$. Note that       
$\phi _{d}$ can be computed from Eq.(\ref{phid}). New Physics 
affecting the moduli of $M_{12}^{\left( q\right) }$ will thus
be seen here, provided that the relative contribution of NP
to each sector is different.

\begin{figure}[t]
\begin{center}
\epsfig{file=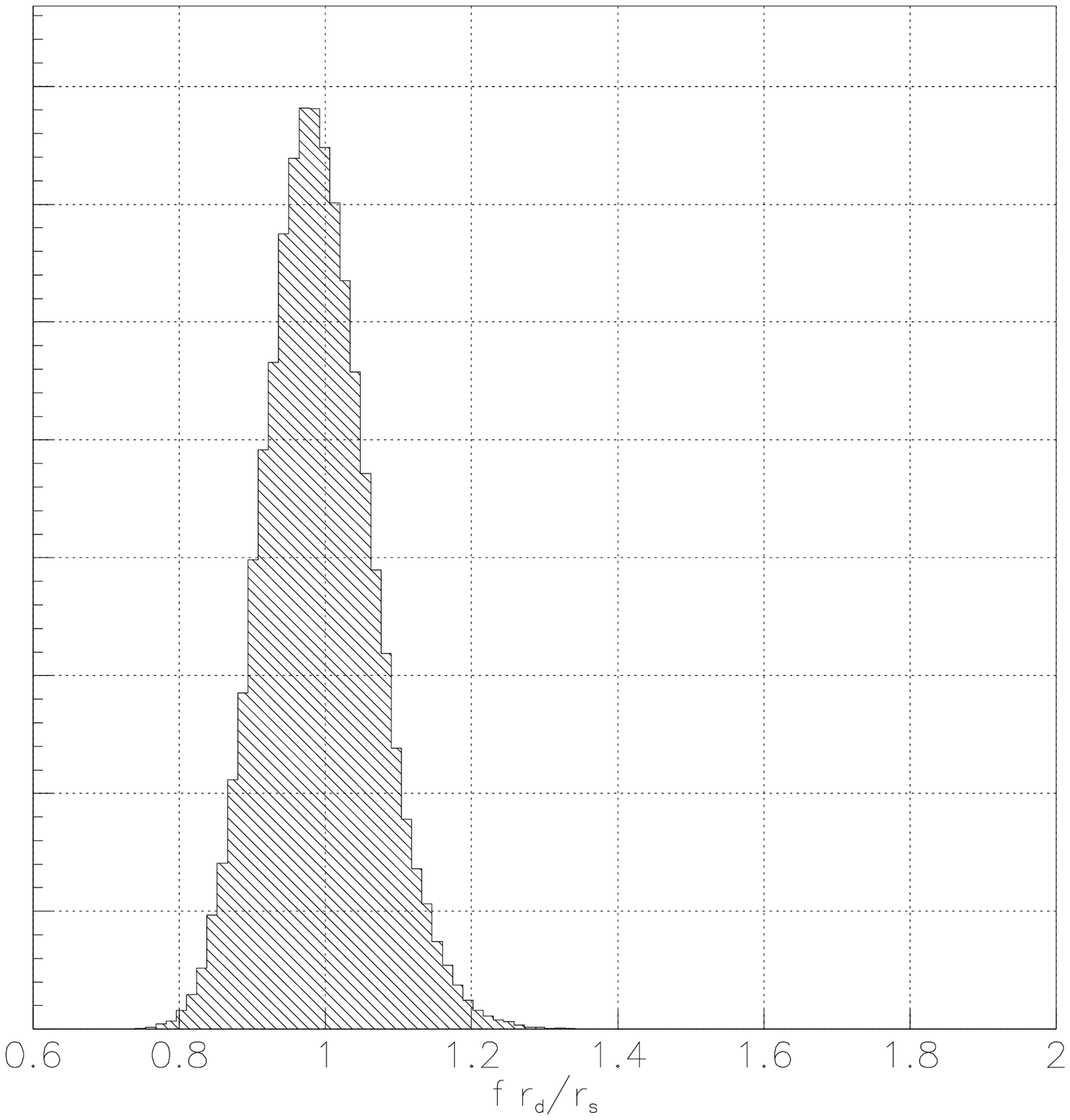,width=3.0493in}
\end{center}
\caption{Distribution for $f\left( \phi _{d}\right) r_{d}/r_{s}$ 
for the input values of Eq.(\ref{fuval}) and 
$x_{d}/x_{s}=2.63\times 10^{-2}$, together with a Gaussian 
distribution for  $\xi =1.16\pm 0.05$ }
\label{figura4}
\end{figure}

\begin{figure}[h]
\begin{center}
\epsfig{file=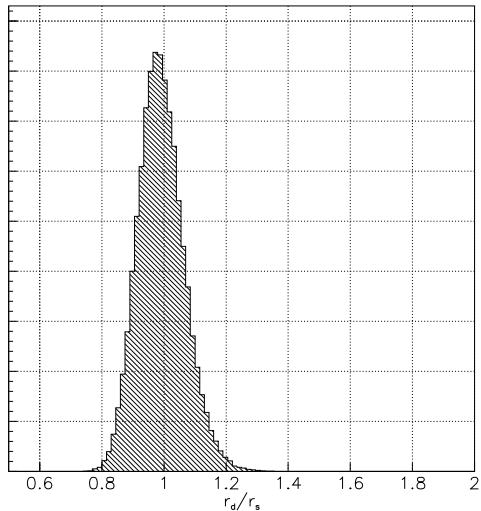,width=3.0493in}
\end{center}
\caption{Distribution for $r_{d}/r_{s}$ 
for the input values of Eq.(\ref{fuval}) and 
$x_{d}/x_{s}=2.63\times 10^{-2}$, together with a Gaussian 
distribution for  $\xi =1.16\pm 0.05$ }
\label{figura5}
\end{figure}

\begin{figure}[h]
\begin{center}
\epsfig{file=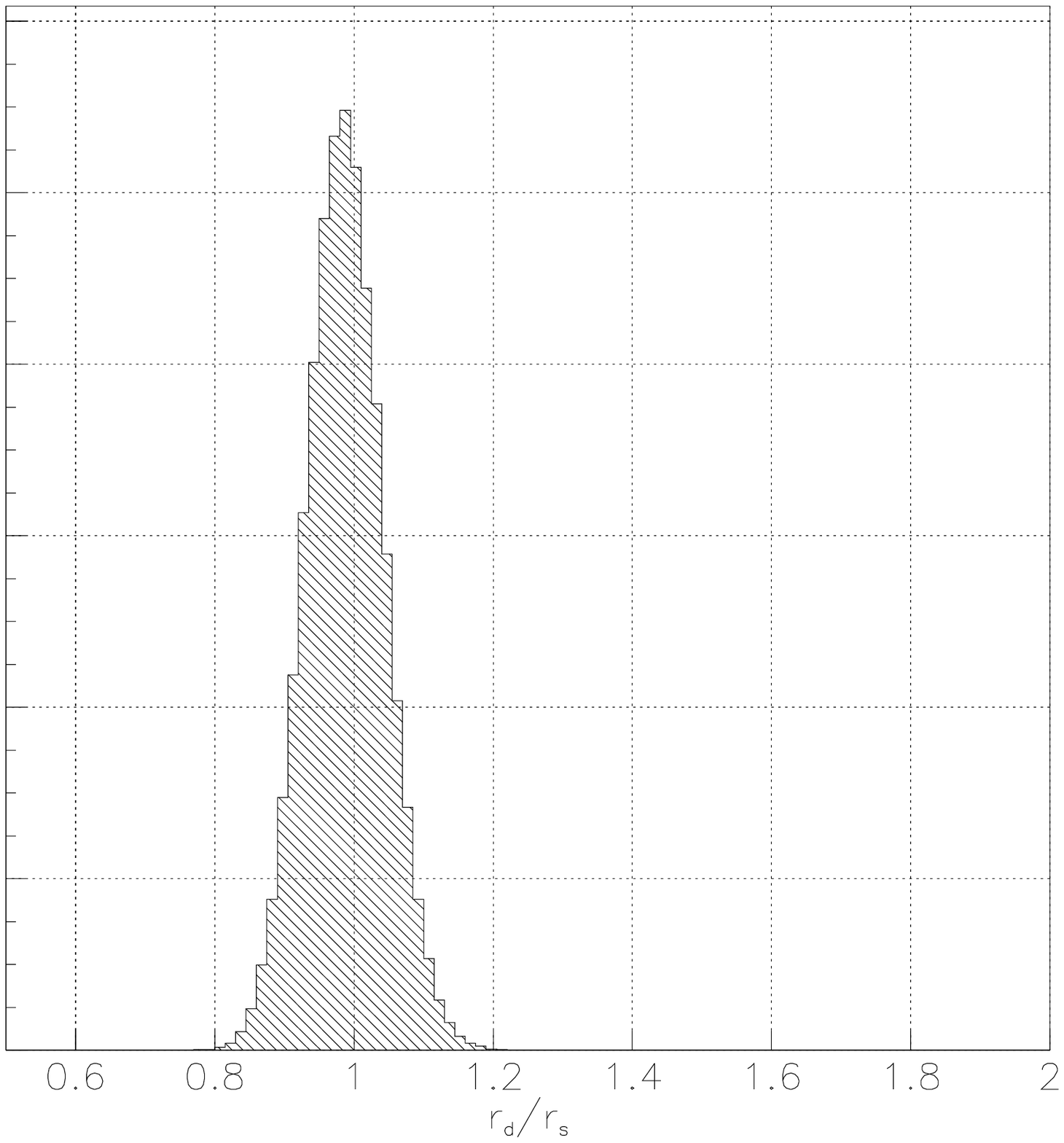,width=3.0493in}
\end{center}
\caption{Distribution for $r_{d}/r_{s}$ for
$\gamma =\left( 56.6\pm 2.6\right) {{}^o}$, 
and the other input values equal to those of Eq.(\ref{fuval}),  
together with $x_{d}/x_{s}=2.63\times 10^{-2}$ and a Gaussian 
distribution for  $\xi =1.16\pm 0.05$ }
\label{figura6}
\end{figure}

By taking, as experimental input the same values as in Example 2,
given by Eq.(\ref{fuval}), together with\footnote{For $x_{s}$ 
we have taken the central value extracted from the other central 
values. We have not included any error for $x_{s}/x_{d}$ consistent 
with BTeV design since it will be too small to play any r\^ ole.} 
$x_{d}/x_{s}=2.63\times 10^{-2}$ and a Gaussian 
distribution for $\xi =1.16\pm 0.05$, we have
generated the toy Monte Carlo distributions for $f\left( \phi _{d}\right)
r_{d}/r_{s}$ and $r_{d}/r_{s}$ presented in Figs.\ref{figura4} and 
\ref{figura5}, respectively. 
%\FRAME{ftbpF}{3.0493in}{3.0493in}{0pt}{}{\Qlb{figura4}}{%
%frdrs-3.eps}{\special{language "Scientific Word";type
%"GRAPHIC";maintain-aspect-ratio TRUE;display "USEDEF";valid_file "F";width
%3.0493in;height 3.0493in;depth 0pt;original-width 7.8793in;original-height
%7.8793in;cropleft "0";croptop "1";cropright "1";cropbottom "0";filename
%'global2/frdrs-3.eps';file-properties "XNPEU";}}
%\FRAME{ftbpF}{3.0493in}{%
%3.0493in}{0pt}{}{\Qlb{figura5}}{rdrs-3.eps}{\special{language "Scientific
%Word";type "GRAPHIC";maintain-aspect-ratio TRUE;display "USEDEF";valid_file
%"F";width 3.0493in;height 3.0493in;depth 0pt;original-width
%7.8793in;original-height 7.8793in;cropleft "0";croptop "1";cropright
%"1";cropbottom "0";filename 'global2/rdrs-3.eps';file-properties "XNPEU";}}%
These distributions are fitted by Gaussians
corresponding to $f\left( \phi
_{d}\right) r_{d}/r_{s}=0.99\pm 0.07$ and $r_{d}/r_{s}=0.99\pm 0.07$
and are almost identical. This is due to the fact that
the  $f\left( \phi_{d}\right) $ distribution resulting from
Eq.(\ref{f}) is quite narrow for the $\phi_{d}$ distribuition
obtained in this example and represented in Fig.\ref{figura2}.
In this case the precision in the  $r_{d}/r_{s}$
determination is limited to $7\%$, due to the error of  $10\%$
assumed for $\gamma $ rather than the error assumed for  $\xi $,
which is of the order of $5\%$. Once $\overline \beta$ and  $x_{s}$
are measured to the accuracy chosen in our example
together with  $\gamma $ to a precision of a few per cent, 
the extraction of 
$r_{d}/r_{s}$ will then be limited by the theoretical knowledge
of  $\xi $. In  Fig. \ref{figura6} we plot the corresponding 
$r_{d}/r_{s}$ distribution for $\gamma =\left( 56.6\pm 2.6 \right)
%TCIMACRO{\UNICODE[m]{0xba}}%
%BeginExpansion
{{}^o}%
%EndExpansion
$, together with the other input values assumed for Fig.\ref{figura5}.
%\FRAME{ftbpF}{3.0493in}{3.0493in}{0pt}{}{\Qlb{figura6}}{rdrs-4.eps}{%
%\special{language "Scientific Word";type "GRAPHIC";maintain-aspect-ratio
%TRUE;display "USEDEF";valid_file "F";width 3.0493in;height 3.0493in;depth
%0pt;original-width 7.8793in;original-height 7.8793in;cropleft "0";croptop
%"1";cropright "1";cropbottom "0";filename
%'global2/rdrs-4.eps';file-properties "XNPEU";}}
In this case, $r_{d}/r_{s}$ is
very well fitted by a Gaussian corresponding to $r_{d}/r_{s}=0.99\pm 0.05$.

\section{Discussion and Conclusions}
We have presented a set of exact relations among specific moduli
and rephasing invariant phases of $V_{CKM}$, which result from
$3 \times 3$ unitarity of the quark mixing matrix. These exact 
relations provide a stringent precision test of the SM, with the 
potential for revealing New Physics. This is specially true if,
on the one hand, $\gamma$, $x_s$ and eventually $\chi$ are
measured in the present or future B factories and, on the other 
hand, there is significant decrease in the theoretical
uncertainties in the evaluation of the relevant hadronic
matrix elements. These tests may complement the standard 
analysis in the $\rho$, $\eta$ plane, which consists of
finding a region in that plane where all experimental
data are accommodated by the SM. Discovering NP corresponds
to having no region in the $\rho$, $\eta$ plane where all data is
accounted for. 
In this NP scenario, our exact relations may be
useful in revealing the origin of NP. For example, if there are
new contributions to $\phi_d$ in 
$B_{d}^{0}-\overline{B}_{d}^{0}$ mixing, but no NP contributions
to $B_{s}^{0}-\overline{B}_{s}^{0}$ mixing, Eq.(\ref{rel9})
will be violated, leading in general to $\phi_d \neq 0$
in Eq.(\ref{phid}), while, for example Eq.(\ref{rel19}) 
will still be satisfied. 
Conversely, if there are NP contributions
to $B_{s}^{0}-\overline{B}_{s}^{0}$ mixing leading to  
$\phi_s \neq 0$, $r_s \neq 1$, but no NP contributing
to $B_{d}^{0}-\overline{B}_{d}^{0}$ mixing, Eq.(\ref{rel9}) 
will hold but Eqs.(\ref{rel17}),(\ref{rel18})and 
(\ref{rel19}) will be violated.

In conclusion, the advent of B-factories offers the possibility
of using exact relations among moduli and rephasing invariant phases
of $V_{CKM}$ to perform precision tests of the SM and hopefully
to uncover the presence of New Physics.

\section*{Acknowledgments}
We would like to thank Helen Quinn and Jo\~ ao Silva for useful
conversations. FJB aknowledges the warm hospitality extended
to him during his stay at IST, Lisbon where part of this 
work was done. FJB received partial support from CICYT AEN-99/0692 and 
from FCT through  Project CERN/P/FIS/40134/2000. GCB and MNR also
received partial support from this project as well as from FCT through 
Project CERN/FIS/43793/2001 and Project CERN/C/FIS/40139/2000
and from the EC under contract HPRN-CT-2000-00148. MN aknowledges
MCED for a research fellowship.

\end{document}